\documentstyle[preprint,aps]{revtex}
\begin{document}
\draft
\title{ $\eta$-Deuteron  Scattering}
\author{A.M. Green\thanks{e-mail "green@phcu.helsinki.fi"}}
\address{ Research Institute for Theoretical Physics, P.O. Box 9,
FIN--00014 University of Helsinki, Finland}
\author{J.A. Niskanen\thanks{e-mail "janiskanen@phcu.helsinki.fi"}}
\address{Department of Physics, P.O. Box 9, FIN--00014 University
of Helsinki, Finland}
\author{S. Wycech\thanks{e-mail "wycech@fuw.edu.pl"}}
\address{Soltan Institute for Nuclear Studies,
Warsaw, Poland}

\maketitle
\begin{abstract}
Eta-deuteron scattering lengths are calculated. A summation of the
multiple scattering series is carried out and the result is checked 
against more  involved calculations. The necessity to go beyond 
the fixed nucleon approximation is emphasized.
It is shown that a quasibound or virtual state in the $\eta$-deuteron system 
may occur within the range of $\eta$-nucleon scattering lengths suggested by 
other experiments. 
\end{abstract}
\pacs{PACS numbers: 13.75.-n, 25.80.-e, 25.40.Ve}

\newpage
\section{Introduction}
\label{intro}
Few-body interactions of the $\eta$ meson may complement our
knowledge on the $\eta$-nucleon interaction. Of related interest is
the possibility of $\eta$-nuclear quasi-bound states. Such states
have been predicted by Haider and Liu \cite{hai} and Li {\it et al.}
\cite{li}, when it was realised that the $\eta$-nucleon interaction
is attractive. In few-nucleon systems these states are expected
to be narrow, and thus easier to detect. So far there has been no
direct experimental verification of this hypothesis.  On the other
hand, Wilkin \cite{wil} has suggested that an indirect effect of such
a state is seen in the rapid slope of the 
$ pd \rightarrow \eta ^{3}$He
amplitude detected just above the $\eta$ production threshold
\cite{gar}. Another indication of the strong three-body $\eta pp$
correlations follows from the recent measurement of $ pp
\rightarrow pp \eta$ cross sections in the threshold region
\cite{cal}.

The $\eta$-deuteron is the easiest few-body system to describe. 
In this letter a simple formula is given to provide the $\eta d$
scattering matrix at energies below the deuteron breakup. Detailed
calculations are done for the scattering length $A_{\eta d}$, where
the $\eta$-nucleon scattering length $a_{\eta N}$ is considered as an
input. In the limiting case of fixed nucleons, this formula is found
to be consistent with earlier calculations of Ref. \cite{rak}. 
However, our model includes corrections involving effects of 
the continuum in the $\eta pp $ system, which are found to be 
necessary. 

For large values of Re $a_{\eta N}$ in the region of 0.7 to 1.0 fm,
suggested by some models, we find the $\eta d$ system to be close to
binding. In this region $A_{\eta d}$ becomes large and depends on
details of the $\eta N$ interaction model. In particular one finds
strong dependence on the way the $\eta N$ scattering matrix is
extrapolated to the region below the threshold. If the actual $A_{\eta
d}$ turns out to be sizable, it will be detected from the analysis 
of the final state interactions in the $ pd \rightarrow pd \eta$ 
scattering experiment performed at Celsius recently \cite{cel}.

\section{A Formula for the eta-deuteron scattering matrix}
\label{sec2}
The purpose of this section is to derive a simple formula
that relates the low-energy meson-deuteron scattering amplitude to the
meson-nucleon scattering length. The former is found by summation
of a multiple scattering series and is expressed in terms of a few
basic multiple scattering integrals. In order to motivate
the method we recall a simple formula for the
scattering length of a meson on a pair of fixed nucleons
\cite{bru,wal}
\begin{equation}
\label{f1}
 A_{\eta d}=
\frac{2a_{\eta N} \xi}{1-\frac{a_{\eta N}}{R_d} \xi} ,
\end{equation}
where $a_{\eta N}$ is the meson-nucleon scattering length and $R_d$
is the nucleon-nucleon distance. Eq. (\ref{f1}) is obtained in 
a simple way by setting a boundary condition for the meson wave 
function $\psi$ at each scatterer $\psi '/\psi = 1/a_{\eta N}$. 
In the simplest version of  genuinely
fixed scatterers $\xi=1$, but a simple correction 
$\xi = m_{\eta d}/\mu_{\eta N}$ is easy to implement.
Here, the meson-deuteron reduced mass $m_{\eta d}$ corrects for the
meson propagator ($1/R_d$), which has to be referred to the $NN$
centre of mass system. The reduced meson-nucleon mass $\mu_{\eta N}$
is necessary to relate the meson-nucleon scattering lengths to the  
meson-nucleon potentials.

Already at this stage Eq. (\ref{f1}) is a fair representation of
$A_{\eta d}$ for Re $a_{\eta N}$ of about 0.3 fm or less. In principle it 
handles also situations of $A_{\eta d} \rightarrow \infty$, i.e. the cases 
of meson-deuteron bound or virtual states close to threshold. The latter 
may occur already at Re $a_{\eta N} \approx 1$ fm, 
which is close to the range of the $\eta$-nucleon scattering 
lengths allowed by some models \cite{bat}. 
However, for such large Re $a_{\eta N}$ Eq. (\ref{f1}) becomes 
rather inaccurate. 
In the rest of this section we find necessary corrections, determine the 
virtual or quasibound state singularities and discuss other related 
calculations of $A_{\eta d}$.

Let us begin with a multiple scattering expansion that follows from the 
three-body Faddeev equations for a meson interacting with a pair of nucleons 
labeled 1,2. For the situation of meson-deuteron scattering below the 
deuteron breakup, the series for the $\eta$-deuteron $T$-matrix is:
\begin{eqnarray}
\label{f2}
T_{\eta d} & = & t_1+t_2+t_1G_0t_2+t_2G_0t_1+t_2G_0t_1G_0t_2+t_1G_0t_2G_0t_1 \nonumber \\
& + & (t_1+t_2)G_{NN}(t_1+t_2)+(t_1+t_2)G_{NN}(t_1+t_2)G_{NN}
(t_1+t_2)+...\, ,
\end{eqnarray}
where $t_i$ is a meson-nucleon  scattering matrix, $G_0$ is
the free three-body propagator and $G_{NN} = G_0 T_{NN} G_0$
is that part of
the three-body propagator which contains the nucleon-nucleon
scattering matrix $T_{NN}$. This expansion is
performed in momentum space and appropriate integrations
over the intermediate momenta are understood.
The detailed notation and normalisation will be given later.
Now the partial summation of the series for the scattering
amplitude is performed. The latter is determined by an average
\begin{equation}
\label{f3} 
<\phi_d \psi_{\eta} \mid T_{\eta d}(E) \mid \phi_d \psi_{\eta}>
\equiv <T_{\eta d}(E)>,
\end{equation}
where $\phi_d$  is the deuteron and $\psi_\eta$ is the
free-meson $s$-wave function
in the relative meson-deuteron momentum variables.
We define the $\eta d$ scattering length as
\begin{equation} 
\label{f4} 
A_{\eta d} = - (2\pi)^2 m_{\eta d} < T_{\eta d}(0) >.
\end{equation}
A partial summation of the series~(\ref{f2}) for $<T>$ is obtained by
\begin{equation}
\label{f5} 
<T^1_{\eta d}>=\frac{<T^0_{\eta d}>}{1-\Omega_1-\Sigma_1}
\end{equation}
where $< T^0_{\eta d} > = < t_1 + t_2 >$ is the impulse
approximation and
\begin{equation}
\label{f6} 
\Omega_1 = \frac{<t_1G_0t_2+t_2G_0t_1>}{<T^0_{\eta d}>} 
\end{equation} 
\begin{equation}
\label{f6a} 
 \Sigma_1  =  \frac{<T^0_{\eta d}G_{NN}T^0_{\eta d}>}{<T^0_{\eta d}>} .
\end{equation}
This partial sum is already equivalent to formula (\ref{f1}) as it contains
terms of the order $\lambda=<t>/R_d$ in the denominator. The expansion
parameter $\lambda$ does not need to be small to guarantee the success of
Eq. (\ref{f5}), which works even for $|\lambda| > 1$, when the multiple
scattering is divergent.
In the $\eta d$ case $|\lambda|$ falls into the 0.1--0.3 range.
Corrections for higher orders of $\lambda$  in the denominator of
Eq. (\ref{f5}) may be obtained by comparing higher orders in 
Eq. (\ref{f2})
with a series expansion of Eq. (\ref{f5}) with respect to
$\Sigma_1$ and  $\Omega_1$. In this way the next approximation is 
obtained
\begin{equation}
\label{f7} 
<T^2_{\eta d}>=\frac{<T^0_{\eta d}>}{1-\Omega_1-\Sigma_1 -
[\Omega_2-(\Omega_1)^2]-[\Sigma_2-(\Sigma_1)^2]
-[\Delta_2-\Omega_1\Sigma_1]}
\end{equation}
where
\begin{equation}
\label{f8} 
\Omega_2=\frac{<t_1G_0t_2G_0t_1+t_2G_0t_1G_0t_2 >}{<T^0_{\eta d}>},
\end{equation}
\begin{equation}
\label{f9} 
 \Sigma_2=\frac{<T^0_{\eta d}G_{NN}T^0_{\eta d}G_{NN}T^0_{\eta d}>}
{<T^0_{\eta d}>} ,
\end{equation}
and
\begin{equation}
\label{f9a} 
\Delta_2=2\frac{<T^0_{\eta d}G_{NN}(t_1G_0t_2+t_2G_0t_1)>}
{<T^0_{\eta d}>} .
\end{equation}

This procedure may be continued into a systematic method to include higher 
powers of $\lambda$ in the denominator. The main advantage is a strong 
cancellation in the $\Sigma_2 - (\Sigma_1)^2$ term and also higher 
order $\Sigma_n$ terms, as was demonstrated in the optical model calculations of
Ref. \cite{hel} for He nuclei. As we show numerically in the next section, 
also for the $\eta$-deuteron system there is a strong cancellation in the 
second (and indeed also in the higher orders) term of the partial sum 
(\ref{f7}). This causes the method to converge much more rapidly than the
direct $\lambda^n$ series.

Before going further we write down and discuss the basic
quantities entering this formalism. Momentum variables
are used everywhere. These are the momenta canonical to
the Jacobi coordinates, 
$\vec q_{NN}$ the relative $NN$ momentum, $\vec p_\eta$
the relative $\eta-NN$ momentum, and the corresponding variables
for the other possible pairs like ($\vec q_{\eta N}, \;\vec p_N$).
The normalisation is chosen so that a "volume" element is
$d\vec p d\vec q$, the propagator $G_0=(E - E_{NN}(q) - E_\eta(p))^{-1}$ 
and the scattering matrices are
$t_{\eta N1}(q_{\eta N1},q'_{\eta N1},E-E(p_{N2})) \delta
(\vec p_{N2}-\vec p'_{N2})$, i.e. they conserve the spectator
momentum. The $t_{\eta N}$ are normalised in such a way that
\begin{equation}
\label{f10} 
t_{\eta N}(0,0,0)= - \frac{a_{\eta N}}{(2\pi)^2\mu_{\eta N}}
\end{equation}
with the standard convention  Im $a_{\eta N} \geq 0$. Later, 
a separable form $t_{\eta N} = 
v_\eta(q_{\eta N}) a_{\eta N}(E) v_\eta(q'_{\eta N}) $
is used with a Yamaguchi formfactor
$v_\eta = (1 + q_{\eta N}^2 / \kappa^2_\eta )^{-1}$.
The $NN$ scattering matrix
$T_{NN} (q_{NN}, q'_{NN}, E-E_\eta(p_\eta))$
is normalised with a different (standard) sign convention that requires
\begin{equation}
\label{f11} 
T_{NN}(0,0,0) = \frac{a_{NN}}{(2\pi)^2 \mu_{NN}}.
\end{equation}
Also here a Yamaguchi separable form is used with 
$\kappa_{NN}= 1.41 $ fm$^{-1}$ 
and the strength fitted to reproduce the deuteron binding energy,
the scattering length $a_{NN}= 5.405 $ fm. 

Calculation of the multiple scattering integrals is straightforward although 
tedious and the formulae are lengthy. For simplicity we reproduce a few 
dominant quantities in the zero meson momentum and zero range
$\eta$-nucleon force limit, although actual calculations are
performed also taking into account a finite force range.
Then the impulse approximation   term becomes
$<T^0_{\eta d} > =- 2 \bar{a}_{\eta N}/( (2\pi)^2\mu_{\eta N})$, where
\begin{equation}
\label{f12} 
\bar{a}_{\eta N}=\int d\vec{p}\, a_{\eta N}(E-\frac{p^2}{2m_{N,\eta N}})\mid
\tilde{\phi}_d(q)\mid^2 
\end{equation}
is the scattering matrix  averaged over some energy region, generated 
by the recoil of the spectator nucleon. The range of the latter is given 
by the Fourier transform of the deuteron wave function $\tilde{\phi}$. 
In a more general nonzero meson momentum case the average is given by 
the momentum distribution of the $\eta N$ pair.

The quantity of interest is the scattering length at threshold 
$A_{\eta d}$. Hence, $E=-E_d$, where $E_d$ is the deuteron binding 
energy, and the energies in Eq. (\ref{f12}) extend down to
the subthreshold region. This means an extrapolation into the
unphysical region by a few MeV. Therefore,
a model is required for this extrapolation and
some possibilities are discussed later. In general,
due to the short range of $\eta N$ forces, the absence of
nearby singularities or an $\eta N$
quasi-bound state, the energy dependence of $a_{\eta N}(E)$
in the narrow subthreshold region is apparently smooth.
In all multiple scattering integrals of interest
the average value $\bar a$ is used. In this way the
dominant term $\Sigma_1$ becomes
\begin{equation}
\label{f13} 
\Sigma_1=-2\bar{a}_{\eta N} \int \frac{d\vec{p}}{(2\pi)^2\mu_{\eta N}}
T_{NN}(E-\frac{p^2}{2m_{\eta d}})[F(p)]^2 \equiv 2 \bar{a}_{\eta N}
\, \sigma_1
\end{equation}
with
\begin{equation}
\label{f14} 
F(p)=\int d\vec{q}\frac{\tilde{\phi}_d(q) \,
v_{NN}(\vec{q}-\frac{1}{2}\vec{p})}
{E- E_{NN}(\vec{q}-\frac{1}{2}\vec{p})-E_{\eta}(p)}\, ,
\end{equation}
where $v_{NN}$ is the Yamaguchi formfactor for the $NN$ separable 
potential. In the low energy region the $T_{NN}$ matrix is dominated 
by the deuteron pole. For $E=-E_d$, the integration in 
Eq. (\ref{f13}) extends from the pole down to negative energies.
When $T_{NN}$ is limited to the pole term, and the $NN$ recoil energy 
$E_{NN}$ is neglected in Eq. (\ref{f14}), expression (\ref{f13}) 
reduces  to a simple form
\begin{equation}
\label{f15} 
  \sigma_1 \approx \frac{m_{\eta d}}{\mu_{\eta N}}
\int \int  d\vec{r} d\vec{r}' \phi^2_d(r) \frac{1}{\mid \vec{r}-\vec{r}'\mid}
\phi^2_d(r') 
\end{equation}
with a clear physical interpretation. The higher order terms for 
$\Sigma_n$ have the same structure corresponding to $(n+1)$
scatterings on the optical potential    
$V_{\eta d} =-2 a_{\eta N} \phi^2_d(r) /(2\pi\mu_{\eta N})$
at zero incident energy. 

Similarly to
Refs. \cite{hel,anti} it can be expected that the series for 
$T^0_{\eta d},\; T^1_{\eta d},\; T^2_{\eta d},...$ converges 
so rapidly that $T^2_{\eta d}$ would be precise on the 1\% level 
even in the case of a bound state at threshold. Indeed, the 
effectiveness of this expansion is confirmed in Table I for 
two particular values of the $\eta N$ scattering lengths 
and an $\eta N$ formfactor  
allowing comparison with the calculation of  \cite{rak}. The 
latter one uses a rather involved set of integral equations for the 
$\eta$ scattering on fixed nucleons corrected later for the effect of 
the deuteron wave functions, but no allowance is made for a 
free $NN\eta$ spectrum in the intermediate states. This assumption
would correspond to our model with $\Omega_i = 0$. The agreement 
between these two calculations is rather good, with small 
differences probably due to two factors: first, 
Gaussian wave functions are used in Ref. \cite{rak},
 while ours come naturally in the Hulthen form; 
second, the $T_{NN}$ used here contains 
more than the deuteron pole. 

The effect of the free three-body spectrum in the intermediate states is 
still missing. To lowest order in $a_{\eta N}$ it is given by the 
$\Omega_1$ term of Eq. (\ref{f6}). Within the  average $\bar{a}_{\eta N}$
approximation it becomes
\begin{equation}
\label{f16} 
 \Omega_1=\bar{a}_{\eta N}\int \int \frac{d\vec{q}' d\vec{q}}
{(2\pi)^2\mu_{\eta N}}\frac {\tilde{\phi}_d(\vec{q})\tilde{\phi}_d(-\vec{q}')}
{E_{NN}(\frac{\vec{q}-\vec{q}'}{2})
+E_{\eta}(\vec{q}+\vec{q}')-E} \equiv \bar{a}_{\eta N}\,\omega_1 .
\end{equation}
This quantity is real below the deuteron breakup, which is the region of 
our interest. In a similar way one obtains higher order terms $\Omega_2$ 
etc., which are also real in this region.

Now, the final formula to be used in the applications is presented as
\begin{equation}
\label{f17} 
A_{\eta d}=2\frac{m_{\eta d}}{\mu_{\eta N}}
 \frac{\bar{a}_{\eta N}}
{1-\bar{a}_{\eta N}(\omega_1+2\sigma_1)-\bar{a}_{\eta N}^2
[(\omega_2-\omega^2_1)+4(\sigma_2-\sigma_1^2)
+4(\delta_2-\omega_1\sigma_1)]},
\end{equation}
where the second order terms are defined in analogy to
Eqs. (\ref{f13}, \ref{f16}), i.e.
$\sigma_2 = \Sigma_2 /(4\bar{a}_{\eta N}^2), \; 
\omega_2 = \Omega_2/\bar{a}_{\eta N}^2$
and $\delta_2 = \Delta_2 /(4\bar{a}_{\eta N}^2)$. The numerical factors 1, 2 
and 4 in Eq. (\ref{f17}) arise from the number of independent 
collisions on the successive nucleons.

Numerical results for the scattering length are given in 
the next section. Before presenting these let us discuss 
the question of the unitarity of $<T_{\eta d}(E)>$, when 
it is calculated for finite meson
energies, but below the deuteron breakup threshold.
Imaginary contributions to $\omega_i$ arise only above the
deuteron breakup, but the absorptive parts of $\sigma_i$
begin already at $E > -E_d$. These are generated by the deuteron pole 
in $T_{NN}(E-E_{\eta}(p))$ in the integral (\ref{f13}) for $\sigma_1$ 
and in similar formulas for the higher order $\sigma_i$. At $E = -E_d$ , 
the pole term of $G_0 T_{NN} G_0$ contributes the 
$\phi_d (r)\phi_d (r')/|\vec r - \vec r'|$ term to expression 
(\ref{f15}) in the coordinate representation. For higher energies 
the meson propagator becomes
$\exp(ip_{\eta d} |\vec r - \vec r'|)/|\vec r - \vec r'|$,
where $p_{\eta d}$ is the incident meson momentum.
Hence for small
momenta we have Im $\sigma_1 \approx ip_{\eta d} m_{\eta d}
/\mu_{\eta N}$. In addition the second term $\sigma_2 -
\sigma_1^2$ generates no terms linear in $p_{\eta d}$ as may be 
seen from Eq. (\ref{f15}) and its second order analog. This allows us 
to present the low energy behaviour of $<T_{\eta d}>$ as 
\begin{equation}
\label{f18} 
<T_{\eta d}(E)> (2\pi)^2 m_{\eta d}=
- [\frac{1}{A_{\eta d}(E)}-ip_{\eta d}]^{-1}
\end{equation}
as required by unitarity. It also permits the use of the series 
summation method in some energy region close to the threshold.

\section{RESULTS}

The formula for the meson-deuteron scattering length
$A_{\eta d}$ expresses it in terms of an "effective
$\eta$-nucleon scattering length" $\bar{a}_{\eta N}$, which is
an average of the $\eta N$ scattering matrix extrapolated
by a few MeV below the threshold. This is our input parameter. 
Inherently there is another parameter, the inverse range 
$\kappa_{\eta N}$ in the $\eta N$ form factor, 
included in the calculated
quantities multiplying powers of the $\bar{a}_{\eta N}$ in   
the expansion (\ref{f17}) for the inverse $1/A_{\eta d}$.
These follow from the three-body interactions, and
they are given in Table II for four representative values
of $\kappa_{\eta N}$ from the literature \cite{rak,bha,ben},
which presumably cover the whole range of allowed values.
It follows from this table that the coefficients of the 
$\bar a^2$ order are very small. The second order terms 
$\sigma_2 - \sigma_1^2$ due to the intermediate $NN$ 
interactions have been found to be small already in the 
optical model calculations of $\eta$He scattering \cite{hel}. 
Higher order terms are negligible as may be
seen from the results of Table I.

As shown in Table I the static nucleon approximation is not adequate 
in the region of large Re $a_{\eta N} $ close to the critical values.
The effect of the $NN$ continuum is significant.  
Some care is needed with calculations of the corresponding scattering 
integrals
in the three-body continuum $\omega_i$. These converge slowly and 
depend strongly on the range of the $\eta N$ interaction. 
In particular, the zero range limit cannot be taken for $\omega_3$ 
and higher orders.
Provided an unphysical value $\kappa_{\eta N} = \infty$
is not used to describe the free-spectrum contributions, the
small weight and the cancellations in $\omega_2 - \omega_1^2$  
reduce effects of the continuum to the $\omega_1$ term. 
A similar cancellation happens in the "mixed term" 
$\delta_2-\omega_1\sigma_1$ 
which, although small, dominates the second order term.
All this allows the simple formula (\ref{f17}) to work very
effectively even under
the demanding condition of a nearby singularity.

In Table III one finds values of $A_{\eta d}$ for a number of
$\eta N$ scattering lengths that follow from several
analyses of the combined $\pi N ,\eta N$ coupled channels
and from $\eta$ photoproduction data.
The resulting $A_{\eta d}$ may vary by an order of
magnitude, reflecting a nearby $\eta d$ quasibound
state that may arise at threshold for Re $a_{\eta N}$
close to 0.8 fm. It is also clear that the results depend
on the extrapolation of $a_{\eta N} (E)$ below the
threshold. No detailed models are available in this
region and two simple approaches have been attempted:\\
(1) a constant $a_{\eta N}$ and\\
(2) a typical low-energy form
$a(E)=a_{\eta N}/(1-iq_{\eta N}a_{\eta N})$ as required by 
unitarity.\\
In the latter case the calculations are done using an imaginary
 $ q_{\eta N}=i\,0.367\,{\rm fm}^{-1}$. This value follows 
from the average value of the subthreshold energy 
argument $(-E_{d}-\frac{p^2}{2m_{N,\eta N}})$ involved in 
Eq. (\ref{f12}), (the recoil energy amounts to some 4 MeV). 
Such an extrapolation reduces the effective values of 
Re~$\bar {a}_{\eta N}$ by $10-20 \% $ as compared to the 
threshold Re~$ a_{\eta N}$ values. The sensitivity on this
extrapolation method is due to the nearby singularity.
This singularity may represent a quasi-bound $\eta d$ state, 
if Re $A_{\eta d}< 0$, or a quasi-virtual $\eta d$ state, 
if Re $A_{\eta d}>0$.

A more systematic study of $A_{\eta d}$ is presented in
Figs. 1 and 2, where plots of Im~$a_{\eta d}$ and
Re~$a_{\eta d}$ are given for two typical values of
Im~$a_{\eta N}$ as functions Re~$a_{\eta N}$.
 These are all calculated for  $\kappa_\eta =3.316$ fm$^{-1}$
 and for the two low energy presentations of $a(E)$.
It is seen that the critical value of Re~$a_{\eta N}$,
when the virtual state is formed at the $\eta d$ threshold
and crosses-over to the quasi-bound state is about 0.8 fm.
That is well within the range of expected values.
If this happens, one may observe very strong effects in the final
state $\eta d$ interactions. These may be seen in 
$ pd \rightarrow pd \eta$ scattering experiments,
if relative  $\eta d$ momenta as small as 10-60 MeV/c are 
measured.

\section{Conclusion}
We have applied a multiple scattering series formalism developed
earlier for the $\eta$-helium system to calculate the $\eta$-deuteron 
scattering length.

The conclusions are as follows:

\noindent a) It is possible to sum effectively the multiple
scattering series for the inverse meson-deuteron scattering matrix
below the breakup threshold. The method has the advantage of
being able to incorporate both the influence of $NN$ scattering
and the free three-body intermediate continuum in a relatively
simple but numerically stable way without resorting to exact
Faddeev equations.

\noindent b) For small values of Re~$a_{\eta N} \leq 0.3$ fm  
the fixed nucleons make a good approximation. For larger values
of Re~$a_{\eta N} $ it is necessary to take into account the 
intermediate three-body 
continuum states. The sensitivity to otherwise small
effects is due to the proximity of an $\eta d$ quasibound state. 
For the same reason the   $\eta d$ scattering length is fairly 
sensitive to the subthreshold extrapolation of the $\eta N$
scattering matrix.

\noindent c) A virtual or quasi-bound $\eta d$ system is likely 
to be formed. For Re~$a_{\eta N} \approx 0.8$ fm it occurs
close to the threshold. Such a situation may be easily detected via 
final state interaction studies.

\acknowledgements

 S.W. is grateful to the Research Institute of Theoretical Physics 
in Helsinki for hospitality and the Finnish Academy Grant 31430
for financial support.

\begin{figure}
\caption{The real and imaginary parts of the $\eta$-deuteron 
scattering length as a function of
Re~$a_{\eta N}$ for the constant extrapolation
of the $a_{\eta N}$ below threshold and for 
$\kappa_{\eta N} =3.316$ fm$^{-1}$.
 The curves are given  for   Im~$a =0.2$ fm (solid) and
 for   Im~$a =0.3$ fm (dashed). }
\end{figure}

\begin{figure}
\caption{The real and imaginary parts of the $\eta$-deuteron 
scattering length as a function of
Re~$a_{\eta,N}$ for the energy-dependent extrapolation
 $a_{\eta N}(E) = a_{\eta N}/(1-iq_{\eta N}a_{\eta N}) $ below 
threshold and for $\kappa_{\eta N} =3.316$ fm$^{-1}$.
 The curves are given for   Im~$a =0.2$ fm (solid) and
 for   Im~$a =0.3$ fm (dashed). }
\end{figure}

\begin{table}
\caption{Convergence of the $A_{\eta d}$ expansion for
two values of $ \bar{a}_{\eta N} = a_{\eta N}$. 
Units are fm and $\kappa_{\eta N}=3.316 {\rm fm}^{-1} $.}

\begin{tabular}{lccl}
${a}_{\eta N}$ & 0.27 + i0.22   &  0.55 + i0.30 &
     \\ \hline
$A_{\eta d}^0 $ & 0.66 + i0.54 & 1.35 + i 0.74 &
 $\sigma_i = 0 \quad \omega_i=0$ \\
$A_{\eta d}^1 $ & 0.66 + i0.83 & 1.56 + i 1.96 &
 $\sigma_1 \neq 0 \quad \omega_i=0$ \\
$A_{\eta d}^2 $ & 0.64 + i0.85 & 1.37 + i 2.14 &
 $\sigma_2 \neq 0 \quad \omega_i=0$ \\
$A_{\eta d}^3 $ & 0.64 + i0.85 & 1.37 + i 2.14 &
 $\sigma_3 \neq 0 \quad \omega_i=0$ \\
Ref. \cite{rak} & 0.65 + i0.85 & 1.38  + i 2.15 &  \\
$A_{\eta d}^2 $ & 0.57 + i0.97 & 0.61  + i 2.73 &
 $\sigma_i \neq 0 \quad \omega_i \neq 0$ \\
\end{tabular}
\label{table1}
\end{table}

\begin{table}
\caption{
The calculated integral quantities
 in the multiple scattering sum for $A_{\eta d}$ as a
function of $\kappa_{\eta N}$.
Units are fm$^{-1}$ for $(\sigma_1,\; \omega_1)$ and fm$^{-2}$
for $(\sigma_2,\;\omega_2)$. }

\begin{tabular}{lcccc}
$\kappa_\eta$ (fm$^{-1}$) &  $\infty$ & 7.617 & 3.316 & 2.357 \\
$\sigma_1$  & 0.4440 & 0.4331 & 0.3966 & 0.3679 \\
$\sigma_2$  & 0.2315 & 0.2245 & 0.1966 & 0.1588 \\
$\omega_1$  & 0.3118 & 0.2958 & 0.2546 & 0.2246 \\
$\omega_2$  & 0.2800 & 0.1870 & 0.1045 & 0.0723 \\
$\delta_2$  & 0.1759 & 0.1620 & 0.1249 & 0.1001 \\
$\sigma_2 -\sigma_1^2$ &
	 0.0180 & 0.0075 &  0.0032 & --0.0039 \\
$\omega_2 -\omega_1^2$ &
		0.1828 & 0.0874 & 0.0027 & --0.0025 \\
$\delta_2 -\omega_1\sigma_1$ &
		0.0374 & 0.0338 & 0.0239 & 0.0174 \\

\end{tabular}
\label{table2}
\end{table}

\begin{table}
\caption{
Eta-deuteron scattering lengths in fm, $\kappa_{\eta N}=3.316 fm^{-1} $.}
\begin{tabular}{lccc}
  &  $a_{\eta N}$ & \multicolumn{2}{c}{$A_{\eta d}$} \\
  \cline{3-4}
  &  & $a(E)=a_{\eta N}$ & $a(E)=a_{\eta N}/(1-iq_{\eta N}a_{\eta N})$ \\
Bhalerao-Liu \cite{bha}      
 & 0.27 + i 0.22 & 0.57 + i 0.97 & 0.64 + i 0.81 \\
"modified B-L" \cite{bha,bat}
 & 0.44 + i 0.30 & 0.63 + i 1.93 & 1.01 + i 1.50 \\
Bennhold-Tanabe \cite{ben}  
  & 0.25 + i 0.16 & 0.66 + i 0.71 & 0.66 + i 0.58\\
"modified B-T" \cite{ben,bat}
 & 0.46 + i 0.29 & 0.72 + i 2.04 & 1.11 + i 1.54\\
Abaev-Nefkens \cite{aba}
 & 0.62 + i 0.30 & 0.36 + i 3.36 & 1.65 + i 2.41\\
Wilkin \cite{wil}      
  & 0.30 + i .30  & 0.39 + i 1.28 & 0.58 + i 1.11\\
 & 0.55 + i 0.30 & 0.61 + i 2.73 & 1.40 + i 1.98\\
Arima \cite{ari}      
   & 0.98 + i 0.37 & --2.75 + i 2.77 & --0.06 + i 6.20\\
Sauerman \cite{sau}    
  & 0.51 + i 0.21 & 1.48 + i 2.31 &  1.65 + i 1.39\\
Batinic \cite{bat}
 & 0.888 + i 0.274 & --2.90 + i 4.12 & 2.37 + i 5.79\\
   & 0.876 + i 0.274 & --2.76 + i 4.24 & 2.42 + i 5.55\\
Tiator \cite{tia} 
 & 0.476 + i 0.279 & 0.81  + i 2.15 & 1.22 + i 1.56\\
Krusche \cite{kru}
 & 0.430 + i 0.394 & 0.14  + i 1.91 & 0.65 + i 1.73\\
   & 0.579 + i 0.399 & --0.13 + i 2.64 & 0.93 + i 2.41\\
   & 0.291 + i 0.360 & 0.17  + i 1.35 & 0.42 + i 1.25\\
\end{tabular}
\label{table3}
\end{table}

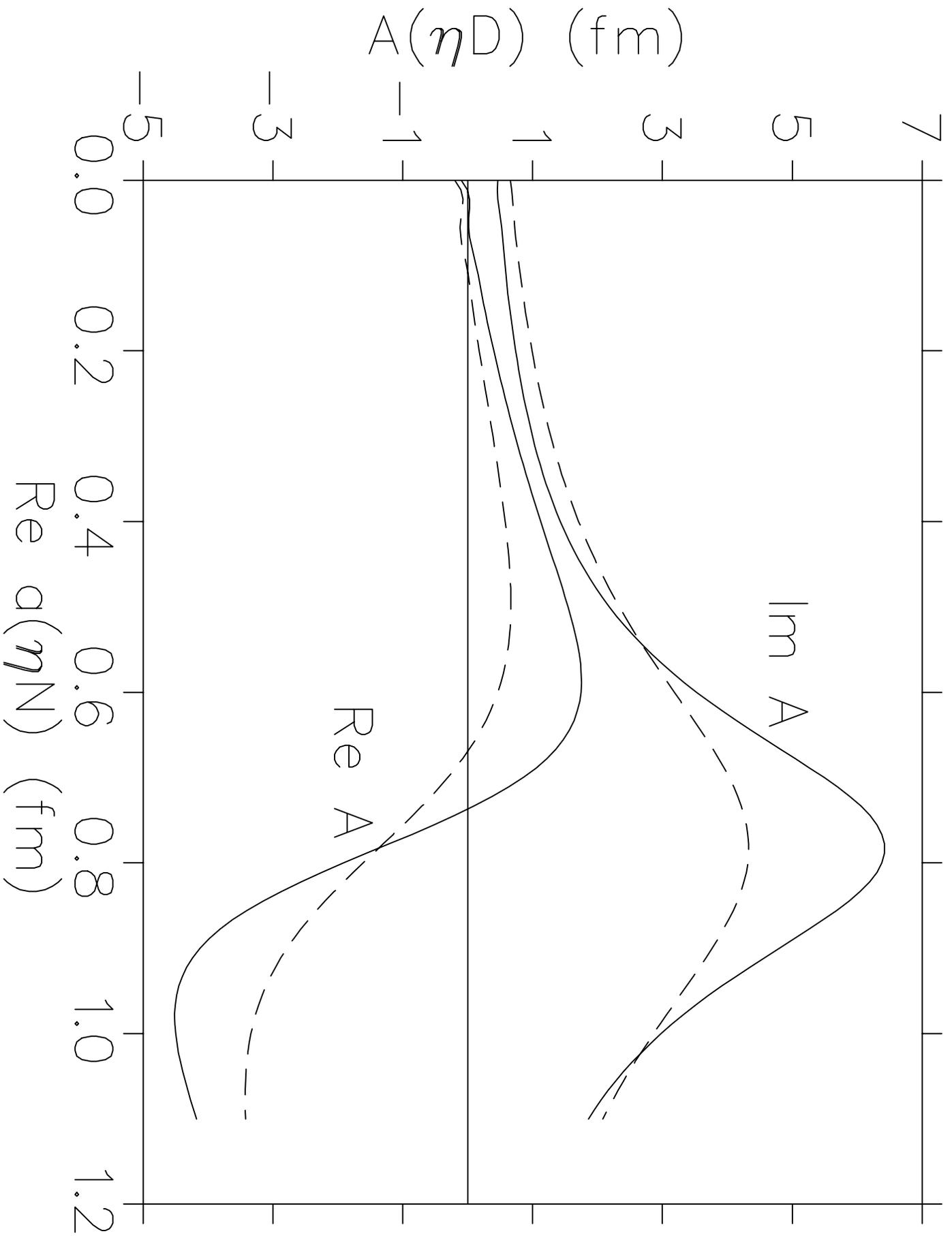
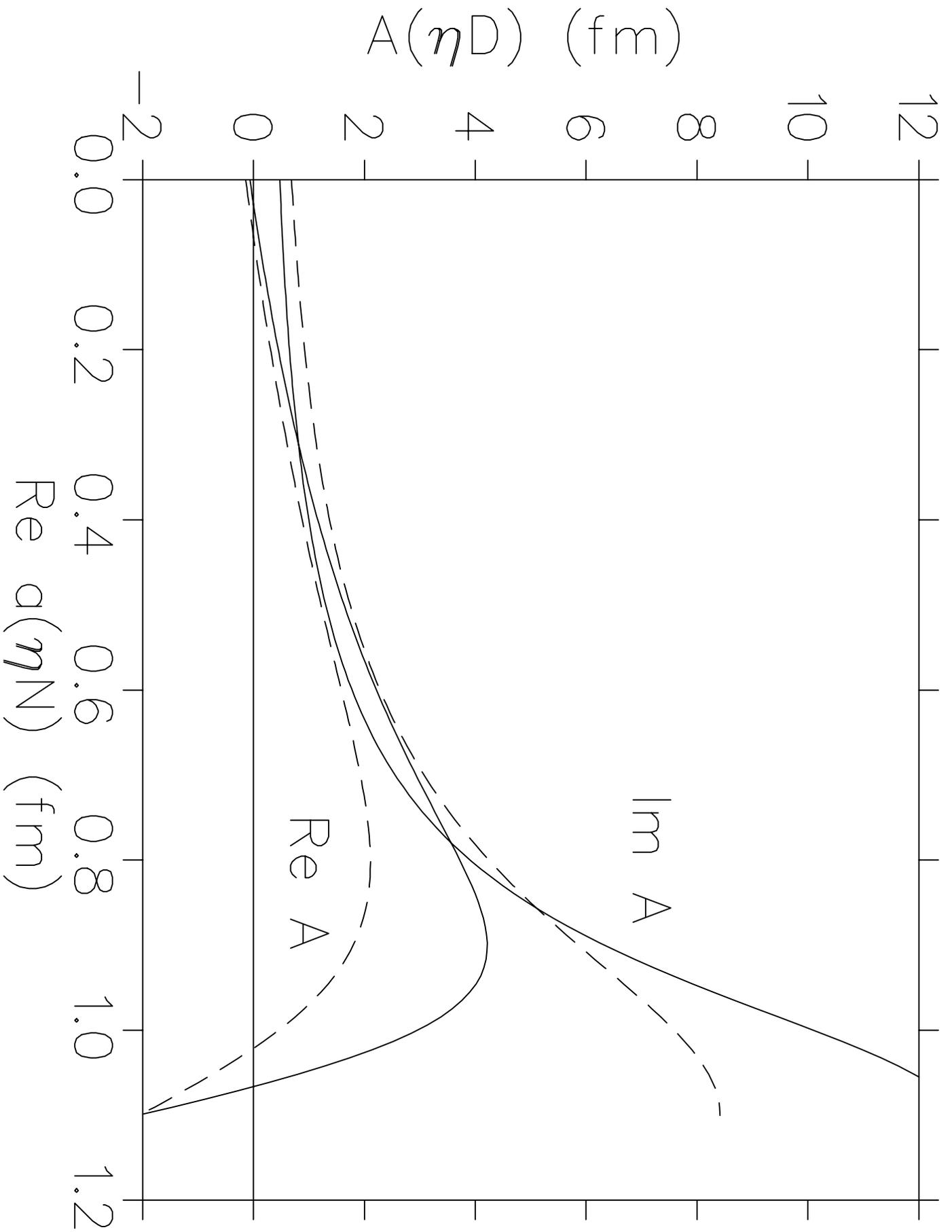
\end{document}